%Paper: 9203055
%From: MARTELLINI%MILANO.INFN.IT@icineca.cineca.it
%Date: Fri, 20 MAR 92 13:17 N

%%%%%%%%%%%%%%%%%%%%%%%%%%%%%%%%%%%%%%%%%%%%%%%%%%%%%%%%%
%
%  INTERACTING EUCLIDEAN 3D QUANTUM GRAVITY
%  G.Bonacina, A.Gamba and M.Martellini
%
%%%%%%%%%%%%%%%%%%%%%%%%%%%%%%%%%%%%%%%%%%%%%%%%%%%%%%%%%

%%%%%%%%%%%%%%%%%%%%%%%%%%%%%%%%%%%%%%%%%
%  use PHYZZX macro package;
%  (figures can be sent upon request)
%%%%%%%%%%%%%%%%%%%%%%%%%%%%%%%%%%%%%%%%%

\def\sqr#1#2{{\vcenter{\hrule height.#2pt
   \hbox{\vrule width.#2pt height#1pt \kern#1pt
   \vrule width.#2pt}
   \hrule height.#2pt}}}

\def\vbsquare{\mathchoice\sqr67\sqr67\sqr{5.1}6\sqr{4.1}6}

\input phyzzx
\PHYSREV

\def\zet{{Z \kern-.45em Z}}
\def\complex{{\kern .1em {\raise .47ex \hbox {$\scriptscriptstyle |$}}
\kern -.4em {\rm C}}}
\def\real{{\vrule height 1.6ex width 0.05em depth 0ex
\kern -0.06em {\rm R}}}
\def\rational{{\kern .1em {\raise .47ex \hbox{$\scripscriptstyle |$}}
\kern -.35em {\rm Q}}}

\input [martellini.subtex]mathsymb
\def\mezzo{{1\over 2}}

\def\widehat{\mathaccent"0362}

\titlepage
\title{Interacting Euclidean Three-Dimensional Quantum Gravity}
\author{ Giuseppe Bonacina}
\address{ Dipartimento di Fisica, Universit\`a di Milano,
I-20133 Milano, Italy}
\author { Andrea Gamba}
\address{Dipartimento di Fisica, Universit\`a di Milano,
I-20133 Milano, Italy}
\andauthor{ Maurizio Martellini}
\address {\it Dipartimento di Fisica, Universit\`a di Milano,
I-20133 Milano, Italy
\break
I.N.F.N., sezione di Pavia, Pavia, Italy}
\vskip .6truein
\abstract
We show that Euclidean 3D-gravity coupled to a Gaussian scalar massive matter
field in first-order dreibein formalism gives a quantum theory which has
a {finite} perturbative expansion around a non-vanishing
background. We also discuss a possible mechanism to generate a non-trivial
background metric starting from Rovelli-Smolin's loop
observables.
\submit{Physical Review D}
\vskip .4truein
\noindent
%CERN--TH.5768
\endpage
\pagenumber=1

\chapter{Introduction}
Some time ago, Witten
\Ref\wittenuno{
E.Witten, Nucl.Phys. {\bf 311B} (1988/89) 46.}
showed that pure 3D-quantum gravity (QG) in first-order dreibein formalism
is a finite off-shell (topological) theory when expanding around a vanishing
background, \ie\ $\langle e^a_{\mu}\rangle =0$.
This result comes from the fact that
Einstein 3D-theory is off-shell and hence it is
dependent on the variables that represent the gravitational field. Later,
Deser et al.
\Ref\deser{
S.Deser, J.McCarthy and Z.Yang, Phys.Lett. {\bf 222B} (1989) 61.}
extended Witten's result showing that the theory remains finite even when
expanding around a flat background gravitational field $\langle e^a_{\mu}
\rangle\propto\delta^a_{\mu}$ (in Euclidean space). Of course, we now have
that $Det\langle e^a_{\mu}\rangle\neq 0$.

In this work, we shall demonstrate that, as far as a perturbative theory
is concerned, Euclidean 3D-gravity coupled to a Gaussian
scalar massive matter field
still yields to a finite quantum theory in first-order dreibein formalism.

In the end, we shall discuss a possible ``quantum''
mechanism for generating as an ``order parameter'' a non-trivial
metric background from some
gauge-invariant and diff-invariant non-local observables
of the pure topological theory, \ie\ of 3D-QG itself. These observables and
their algebra were first introduced by Rovelli and Smolin
\Ref\smolin{
C.Rovelli and L.Smolin, Phys.Rev.Lett. {\bf 61} (1988) 1155; Nucl.Phys.
{\bf 331B} (1990) 80.}
in the frame of Ashtekar's reformulation of canonical 4D-general relativity
\Ref\ashuno{
A.Ashtekar, Phys.Rev.Lett. {\bf 57} (1986) 2244; Phys.Rev. {\bf 36D} (1987)
1587.}
recently specialized to the case of 2+1-dimensional Einstein gravity.
\Ref\ashdue{
A.Ashtekar, V.Husain, C.Rovelli, J.Samuel and L.Smolin, Class.Quantum Grav.
{\bf 6} (1989) L185.}
\endpage

\chapter{Path Integral Formulation of the Theory}
First-order dreibein gravity with Euclidean signature is described by the
action
$$I_E =\int d^3 x\ \varepsilon^{\mu\nu\lambda}e_{\mu a}[\partial_{[\nu}\omega
^a_{\lambda ]}+\varepsilon^a_{bc}\omega^b_\nu\omega^c_\lambda],\eqno (1)$$
where we have absorbed a $k^{-1}$ factor into the dreibein $e^a_\mu$ and
the spin connection $\omega^a_\mu =\varepsilon^{abc}\omega_{\mu bc}$ is an
independent variable. In the following we shall consider the coupling of
Eq. (1) to a real scalar massive matter field which has the first-order
action
\Ref\taylor{
J.G.Taylor, Phys.Rev. {\bf 18D} (1978) 3544.}
$$I_M=\mezzo\int d^3 x\ [\Phi^a \sqrt{e}e^\mu_a\partial_\mu\varphi+\mezzo
(\Phi^a)^2+em^2\varphi^2]$$
where $\Phi$ is a Lagrange multiplier, $e^\mu_a$ is formally the inverse
matrix $[e^a_\mu]^{-1}$, $e\equiv Det^{-1}(e^\mu_a)$ and, of course, we
assume that $Det(e_{\mu a})\neq 0$. Notice that the Euclidean metric
$g_{\mu\nu}$ is given by: $g_{\mu\nu}=e^a_\mu e^b_\nu\delta_{ab}$.
This equation can be rewritten in the form
$$
I_M=\mezzo\int d^3 x\ (\widehat {e}^\mu_a \widehat {e}^{\nu a}\partial_\mu
\varphi\partial_\nu\varphi+ em^2\varphi^2), \eqno (2)
$$
after using the equation of motion of $\Phi$ and setting
$\widehat {e}^\mu_a=\sqrt{e} e^\mu_a$, where
$\widehat {e}^\mu_a$ are tensor densities of weight $1/2$.
Now, we need to fix a gauge.
We choose a Landau-type gauge:
$$\partial^\mu e^a_\mu=0=\partial^\mu\omega^a_\mu. \eqno (3)$$
The resultant ghost and gauge-fixing action is then
$$\eqalign{
I_{FP+GF}=\int d^3 x\ &\{ {\cal C}_a \partial^\mu\omega^a_\mu+{\cal D}_a
\partial^\mu e^a_\mu+\overline{c}_a\partial^\mu[(\partial_\mu\delta^a_b+
\varepsilon^a_{cb}\omega^c_\mu)c^b] \cr
&+ \overline{d}_a\partial^\mu(\varepsilon^{acb}
e_{\mu c}c_b)+\overline {d}_a\partial^\mu[(\partial_\mu\delta^a_b+
\varepsilon^a_{cb}\omega^c_\mu)d^b]\} \cr} \eqno (4)$$
where ${\cal C}_a$ and ${\cal D}_a$ are Lagrange multipliers, $c_a$,
$\overline {c}^b$ and $\overline {d}_a$, $d^b$ are Fadeev-Popov ghosts. The
sum of Eq. (1), Eq. (2) and Eq. (4) gives the total quantum action $I$. The
corresponding Euclidean path integral has the form:
$$\int {\cal D}e^a_\mu{\cal D}\omega^a_\mu{\cal D}\varphi{\cal D}\overline{c}_a
{\cal D}c^b{\cal D}\overline {d}_a{\cal D}d^b\ e^{-I}. \eqno (5)$$

The action $I$ is invariant under the following nilpotent (on-shell)
BRST-transformation~$s$~\Ref\becchi{
G.Gonzales and J.Pullin, Phys.Rev. {\bf 42D} (1990) 3395, {\bf 43D} (1991)
2749;
\nextline
E.Guadagnini, N.Maggiore and S.P.Sorella, Phys.Lett. {\bf 247B} (1990) 543.}
$$\left\{
\eqalign{&s \omega^a_\mu=-(D_\mu c)^a \cr
&s c^a=\mezzo\varepsilon^a_{bc}c^b c^c,\quad s\overline{c}_a={\cal C}_a,\quad
s{\cal C}_a=0 \cr
&s d^a=\varepsilon^a_{bc}c^b d^c,\quad s\overline{d}_a={\cal D}_a,\quad
s{\cal D}_a=0 \cr
&s e^a_\mu=-(D_\mu d)^a-\varepsilon^a_{bc} e^b_\mu c^c \cr
&s \varphi=-e^\mu_a c^a\partial_\mu\varphi \cr
&D_{\mu b}^a \equiv(\partial_\mu\delta^a_b+\varepsilon^a_{cb}\omega^c_\mu) \cr
&s^2 =0 \cr}\right. $$

Our purpose is to integrate out, first, the matter field in the functional
integral. Exact Gaussian integration of the matter field gives:
$$Ne^{-W}\equiv\int{\cal D}\varphi\ e^{-I_M}=N[{\rm Det}(\widehat{\vbsquare}+
\widehat{m}^2)]
^{-\mezzo}=Ne^{-\mezzo {\rm lnDet}(\widehat{\vbsquare}+\widehat{m}^2)},
\eqno (6)$$
where $\widehat {\vbsquare}=-\widehat{e}^\mu_a\partial_\mu \widehat{e}^{\nu a}
\partial_\nu=-\widehat{\partial}_a\widehat{\partial}^a$ and $\widehat{m}^2
=em^2$. Here, $W=\mezzo {\rm lnDet}(\widehat{\vbsquare}+
\widehat{m}^2)$ is the one-loop
effective action of the matter field. Following Birrell and Davies,
\Ref\birrell{
N.D.Birrell and P.C.W.Davies, ``Quantum Fields in Curved Space'', Cambridge
Monographs on Math.Phys. vol.7.}
we see that using the DeWitt-Schwinger representation and dimensional
regularization, $W$ can be written as
$$W_{m^2}=\int d^n x\ {\cal L}_{eff}(x)\equiv\int d^n x\ \sqrt{g(x)} L_{eff}
(x;m^2),$$
where, in $n$ dimensions, the asymptotic (adiabatic) expansion of $L_{eff}$ is
$$L_{eff}\simeq\mezzo(4\pi)^{-{n\over 2}}
\sum^\infty_{j=0}a_j(x)(m^2)^{{n-2j\over
2}}\Gamma(j-{n\over 2}). \eqno (7)$$
We immediately notice that, in \underbar{odd dimensions}, $L_{eff}$ has only
\underbar{finite terms} because $\Gamma(j-{n\over 2})<\infty$ and $a_j(x)$ are
geometrical invariants built out of the curvature tensor and its contractions.
In three dimensions, for \underbar{large external}
momenta (or equivalently for large
$m^2$ in Planck units), only the first two terms of Eq. (7) are sensibly
different from zero.

After the functional integration (6), we are left with the loop expansion
constructed from the effective lowest order action
$$I'\equiv I_E+I_{FP+GF}+W_{m^2}. \eqno (8)$$
In a quantum perturbative treatment of $I'$ (see next chapter) one has to
consider the modified Feynman rules, coming from
$W_{m^2}$, of $I_E+I_{FP+GF}$.
Indeed, the first two terms of $L_{eff}$ may be regarded as a contribution
to the gravitational lagrangian although they arise from the action of the
quantum matter field, since one has that
$a_0(x)=e$ and $a_1(x)={e\over 6} R(e,w)$,
where $R(e,w)$ is the curvature scalar thought in first-order dreibein
formalism (Eq. (1)). Notice that for large $m^2$ (in Planck units),
$I_E(e,w)+W_{m^2}(e,w)$ is equivalent to a Euclidean non-Abelian
Chern-Simons gauge theory with gauge group
$SO(4)\sim SU(2){\rm X}SU(2)$, assuming
that the induced cosmological constant $\Lambda\equiv {m^3\over 12\pi^2}$ is
positive definite [1].
As a consequence,
we may understand the effects of coupling
Gaussian matter fields to 3D-gravity
in the ultraviolet region,
\ie\ for large momentum of the graviton,
as a ``dressing'' of the pure 3D-gravity sector.
Thus, one still ends with a
Chern-Simons gauge theory which, according to
common wisdom [1],
gives a finite quantum theory. This is actually the
case, as we shall see in
the next chapter by arguments of power counting.
\chapter{Perturbative Expansion}
As usual, the exact treatment of Eq. (8) is too hard a thing to cope with
and therefore we go on with a perturbative expansion. To establish finiteness
of Eq. (8) around a flat background $e^a_\mu=\delta^a_\mu+h^a_\mu$, $h<<1$
(in Euclidean space), we keep the gauge of Eq. (3) that we now write
$\partial^\mu h^a_\mu=0=\partial^\mu\omega^a_\mu$. Remembering that we move
indices with $\delta^a_\mu$, \ie\ we identify the metric introduced in the
gauge
fixing with that introduced by the background, we write the resultant
ghost and gauge fixing action as [2]
$$\eqalign{
I_{FP+GF}=\int d^3 x\ &\{ {\cal C}_a\partial^\mu\omega^a_\mu+{\cal D}_a
\partial^\mu h^a_\mu+\overline{c}_a\partial^\mu[(\partial_\mu\delta^a_b+
\varepsilon^a_{cb}\omega^c_\mu)c^b] \cr
&+ \overline{d}_a\partial^\mu[\varepsilon^a_{cb}(\delta^c_\mu+h^c_\mu)c^b+
(\partial_\mu\delta^a_b+\varepsilon^a_{cb}\omega^c_\mu)d^b]\} \cr}. \eqno (9)$$
The vertices of this action and of the pure gravitational one $I_E$ are cubic,
while the three basic (off-diagonal) propagators are [2]
$$\eqalignno{
&\langle \omega^b_\nu h^a_\mu\rangle=-i\delta^{ab}\delta_{\mu\alpha}
\delta_{\nu\beta}\varepsilon^{\alpha\beta\gamma}{p_\gamma\over p^2}
\rightarrow{1\over p},\quad as\ {p\rightarrow\infty} & (10.a) \cr
&\langle \overline{c}_a c^b\rangle={\delta^b_a\over p^2}=\langle\overline{d}_a
d^b\rangle. & (10.b) \cr}$$
In the following we shall represent the graviton propagator
$\langle\omega h\rangle$ as in \FIG\graprop{Graviton propagator $\langle
\omega h\rangle$.} Fig.~\graprop, where the dashed (wavy) line stands for
$\omega$ ($h$).
In addition, we should have propagators due to the flat background but,
for simplicity, we treat them as new vertices beside the cubic ones. We now
have to consider the contribution to the above Feynman rules (Eq. (10)) due to
the matter one-loops in Eq. (8), \ie\ in
$\mezzo {\rm lnDet}\{[\widehat{\vbsquare}+
\widehat{m}^2](\delta +h)\}$,
which shall be treated in the standard perturbative
expansion in the quantum field $h^a_\mu$. This amounts to calculating the
effective graviton propagators and vertices given at the lowest order by
the insertion of matter one-loops in the $h$-lines alone.
In this connection, we
need the $\varphi$-propagator and the $h\varphi\varphi$-cubic vertex. In
this perturbative framework, the two (three)-point function $\langle\varphi
\varphi\rangle$ ($\langle h\varphi\varphi\rangle$) is obtained from Eq. (2)
expanding $e^a_\mu$ up to order $O(h^2)$, \ie\ in the momentum representation
picture one must start from
$$\left\{
\eqalign{&I_M =\int d^3 p\ [\mezzo(p^2+m^2)\varphi^2+s^{\mu\nu}(p_\mu p_\nu-
\delta_{\mu\nu}m^2)\varphi^2] \cr
&s^{\mu\nu} =\delta^\mu_a\delta^{\rho\nu}s^a_\rho,\quad s^a_\rho\equiv
h^a_\rho-\mezzo\delta^a_\rho h,\quad h\equiv\delta^\rho_a h^a_\rho \cr}
\right.. \eqno (11)$$
As a consequence we get the matter Euclidean Feynman rules:
$$\langle\varphi\varphi\rangle={1\over p^2+m^2},\qquad\qquad \langle
\varphi\varphi s^{\alpha\beta}\rangle=-2(p_\alpha p_\beta-\delta_{\alpha\beta}
m^2). \eqno (12)$$
We pictorially associate $\langle\varphi\varphi\rangle$
and $\langle \varphi\varphi s^{\alpha\beta}\rangle$
with \FIG\mattprop{Matter propagator $\langle\varphi\varphi\rangle$.}
Fig.~\mattprop\ and
\FIG\magraver{Matter-graviton vertex $\langle\varphi\varphi s\rangle$.}
Fig.~\magraver\ respectively.
With these Feynman rules, we notice
that the first term coming from
the $h^a_\mu$-expansion of the non-local action $W$ defined in Eq. (6)
gives a constant contribution to the cosmological constant. The other
diagrams, instead, all go like $p^3$ as $p\rightarrow\infty$. In fact,
if we take, \eg, the term quadratic in $h^a_\mu$, the self-energy diagram,
its lowest order contribution to Eq. (10.a) is given by
$$\eqalign{
\int\nolimits_{\omega\rightarrow {3\over 2}}d^{2\omega}k\ &{[k_\alpha(
k+p)_\beta-\delta_{\alpha\beta}m^2][k_\mu(k+p)_\nu-\delta_{\mu\nu}m^2]
\over (k^2+m^2)[(k+p)^2+m^2]} = \pi^3\{{163\over 128}
p_\alpha p_\beta p_\mu p_\nu{1\over \sqrt {p^2}} \cr
&+ ({3\over 128}A_{\alpha\beta\mu\nu\rho\sigma}p^\rho p^\sigma-{5\over 128}
B_{\alpha\beta\mu\nu\eta\lambda}p^\eta p^\lambda-{13\over 12}C_{\alpha\beta
\mu\nu}p^2)\sqrt {p^2}\}+O(m^2) \cr}, \eqno (13)$$
where $A_{\alpha\beta\mu\nu\rho\sigma}=\delta_{\alpha\beta}\delta_{\mu\rho}
\delta_{\nu\sigma}+\delta_{\alpha\nu}\delta_{\rho\mu}\delta_{\sigma\beta}+
\delta_{\beta\mu}\delta_{\rho\alpha}\delta_{\sigma\nu}+\delta_{\mu\nu}
\delta_{\rho\alpha}\delta_{\sigma\beta}$, $B_{\alpha\beta\mu\nu\eta\lambda}=
\delta_{\alpha\mu}\delta_{\eta\beta}\delta_{\lambda\nu}+\delta_{\beta\nu}
\delta_{\eta\alpha}\delta_{\lambda\mu}$ and $C_{\alpha\beta\mu\nu}=
\delta_{\alpha\beta}\delta_{\mu\nu}+\delta_{\beta\mu}\delta_{\alpha\nu}+
\delta_{\alpha\mu}\delta_{\beta\nu}$. Its calculation, using dimensional
regularization, shows that it is finite and it goes like $p^3$ for large
momentum $p\equiv \sqrt{p^2}$. On the other hand, all the graviton vertex
corrections (to $I_E$, Eq. (1)) induced by ${\rm lnDet}(\widehat{\vbsquare}+
\widehat{m}^2)$ in Eq. (8), behave like $p^3$ when the momentum $p\equiv
\sqrt{p^2}$ of one of the gravitons becomes large. Roughly speaking, this
is a consequence of the fact that the matter propagator and vertex grow
as $(1/p^2)$ and $p^2$ for $p\rightarrow\infty$, respectively. This is
the crucial reason why our theory will come out UV finite. Then, the lowest
order effective
graviton propagator and cubic vertex are given respectively in
\FIG\effprop{Effective graviton propagator $\langle\omega s\rangle_{eff}$.}
Fig.~\effprop\ and
\FIG\effvert{Effective vertex propagator $\langle\omega\omega s\rangle_{eff}$.}
Fig.~\effvert. They behave as
$$
\langle\omega^a_\mu s^b_\nu\rangle_{eff}(p)
\sim {1\over p^3},\quad p\rightarrow\infty,  \eqno (14)$$
$$
\langle\omega^a_\mu\omega^b_\nu s^c_\rho\rangle_{eff}(p)
\sim p^3,\quad p\rightarrow\infty. \eqno (15)
$$
Remember that $s^a_\mu\equiv h^a_\mu-\mezzo\delta^a_\mu(\delta^\rho_b
h^b_\rho)$
and $p$ is a Euclidean momentum variable.
The off-diagonality of the full 1PI dressed propagator, Eq. (14), which
comes from a straightforward calculation, may be also explained by the
following simple argument. At tree level off-diagonality comes
from the fact that Euclidean 3D-gravity is a (three dimensional) non-Abelian
gauge theory with a \underbar{non-semisimple}
Lie group as an internal simmetry,
namely $ISO(3)$, and hence it has the Lagrangian structure $eR$.
This is a particular case of the so-called BF-theories.
In our dimensional regularization scheme the above gauge invariance
is preserved in the full dressed effective action, Eq. (8), and
therefore it still has a BF-structure.

The important consequence of Eq. (14)
is that now the graviton fields ($\omega,h$) should be assigned ultraviolet
dimension \underbar{zero} in place of the canonical value one of pure gravity,
while the dimension of the ghost fields is one-half as usual. Taking into
account the above considerations, we find the superficial degree of divergence
$\omega(G)$ of an arbitrary diagram $G$ by standard dimensional power
counting as
$$\omega_{dim}(G)=3+\sum_V(\omega_V-3)-2E_{\overline{d}}, \eqno (16) $$
where $\omega_V$ is the dimension of the interaction monomial attached
to a generic vertex appearing in Eq. (8) and $E_{\overline{d}}$ is the
number of external anti-ghosts $\overline{d}$. Notice that in
the power counting (16) the graviton external legs in ($\omega,h$) do not
contribute since $dim_{UV}(h)=dim_{UV}(\omega)=0$.

It is worth mentioning that the
power counting (16) makes also sense for Feynman
graphs with external lines included, which in our approach involve, in
principle, only the
graviton and the ghost fields (although only the ghost fields contribute to
(16)). Indeed, our basic idea in this work is to
treat the Gaussian quantum matter $\phi$-field as a correction to the graviton
self-energy and therefore to the effective graviton propagator, thus obtaining
zero UV-dimension for the graviton field. From a technical point of view
this is always possible, since we can exactly integrate out the boson $\phi$
in the vacuum amplitude functional integral (5).
\foot{
A similar idea was implemented by Tomboulis
\Ref\tomb{
E.Tomboulis, Phys.Lett. {\bf 70B} (1977) 361.}
in the $(1/N)$-expansion of
4D-QG coupled to $N$-massless fermions, and later on extended by
Smolin
\Ref\smol{
L.Smolin, Nuclear Phys. {\bf 208B} (1982) 439.}
to $d$ dimensions.}
As a consequence, our perturbative Feynman rules follow from the effective
action (8), and hence do not contain the $\phi$-field any more.

There is also a formal reason that suggests us that there are no more
renormalizations for the scalar field
$N$-point functions beside those required by
the power counting (16). If at the beginning (\ie\ before the $\phi$-functional
integration) we pick an $N$-point function of the $\phi$-field, after the
introduction of a scalar density source
in the path integral, by using the fact that
the integral is still Gaussian in $\phi$, we find that the scalar $N$-point
function is given by the functional integral with respect to the effective
measure $DeD\omega D(ghosts)e^{-I'}$, with $I'$ defined by Eq. (8), of a
sum, $\sum$, of products of scalar field
propagators. It turns out that $\sum$ may
be factorized out of the functional integral. Hence, we are left only with
the renormalization of the vacuum amplitude functional integral as before.

We now turn to the study of possible divergences. First thing, only
one-loop diagrams can be constructed because of the off-diagonality of the
propagator (Eq. (14)) and the dependence on the field variables of the
vertices [2]. Besides, all the interaction monomials have dimension
$\omega_V=3$ for $V=\langle h\omega\omega\rangle$ and $V=[{\rm lnDet}(
\widehat{\vbsquare}+\widehat{m}^2)-O(h^2)]$ or $\omega_V<3$ for the ghost
vertices ($\omega_{ghosts}=2$).
This tells us that our theory is at least power counting renormalizable.
As a matter of fact, considering, at first, graphs without ghost vertices,
we see that we have a superficial cubic degree of divergence for any number
$n$ of external (graviton) legs. However, they vanish using the dimensional
regularization scheme which implies that $\int{d^{2d}k\over
(2\pi)^{2d}}(k^2)^{\beta-1}=0$ for $\beta=0,1,2,\ldots$ and any $d$
('t Hooft-Veltman conjecture)\
\Ref\leib{
G.Leibbrandt, Rev.Mod.Phys. {\bf 47}(4) (1975) 849.}
and in our case $\beta=1,\ d\rightarrow (3/2)$. While in $D=3$
neither quadratic nor logarithmic divergences are possible for parity reasons,
only linear ones
remain. In this case, diagrams consist of two ghost vertices of type
$\langle\omega c\overline{c}\rangle$ or $\langle\omega d\overline{d}\rangle$.
As we have already noticed, in the dimensional regularization scheme, linearly
divergent graphs are set to zero. Therefore, the quantum theory of 3D-gravity
coupled to a free scalar massive matter field is \underbar{finite} off-shell
in first-order dreibein formalism when it is expanded around a
non-degenerate flat background.

In the end, we should like to observe the following two things. First, we
notice
that the perturbative renormalizability can also be reached in the framework
of a BRST-quantization scheme. Indeed, one could show that the possible
divergent part, $\Gamma_{div}$, of the effective action $I'$, Eq. (8),
satisfies
(in our Landau-like gauge) the Ward identity
$${\cal G}\Gamma_{div}=\Bigl(s_{eff}e^a_\mu {\delta\over \delta e^a_\mu}+
s_{eff}\omega^a_\mu{\delta\over \delta\omega^a_\mu}\Bigr)\Gamma_{div}=0,$$
where ($s_{eff}e^a_\mu$, $s_{eff}\omega^a_\mu$) are the BRST-transformations
that leave the effective action $I'$ invariant. Here we have used the fact
that,
as it has been observed, all 1PI-diagrams containing external ghost lines are
convergent in the dimensional regularization scheme and/or for parity-symmetry
reasons. Therefore, $\Gamma_{div}$ does not depend on the ghosts. Then, this
Ward identity tells us that $\Gamma_{div}$ is a BRST-invariant functional of
($e$, $\omega$) alone. Since the divergent part is local and of dimension
three at most, the only possible form $\Gamma_{div}$ is thus given by the
3D-Einstein action (in first-order dreibein formalism) itself.

As a second thing, we would like to notice that in the above renormalization
discussion we have assumed that the effects of Lorentz anomaly
terms (if any) like
$$I_{LCS}={ik'\over 8\pi^2}\int\nolimits_M\varepsilon^{\mu\nu\rho}
(R_{\mu\nu ab}\Gamma^{ab}_\rho+{2\over 3}\Gamma^b_{\mu a}\Gamma^c_{\nu b}
\Gamma^a_{\rho c}), \eqno (17) $$
in Euclidean signature, may be taken into account by enlarging
\Ref\wittendue{
E.Witten, Nucl.Phys. {\bf 323B} (1989) 113.}
the spin-connection gauge group,
which at the classical Euclidean level is $SO(3)\sim SU(2)$.
In Eq. (17), $\Gamma$ and $R$ are the Levi-Civita connection and the
curvature respectively for the
dreibein field $e^a_\mu$, which is the fundamental variable. $I_{LCS}$
can be interpreted as a CS-term for an $SO(3)\sim SU(2)$ gauge connection
$\Gamma$. Thus, in first-order dreibein formalism, this is equivalent to
starting with the $so(3)\bigoplus so(3)$ Lie algebra-valued connection
$\omega\bigoplus \Gamma$, or equivalently to considering the complex
gauge group $SO(3,\complex)\sim SL(2,\complex)$ as ``internal Lorentz''
symmetry.
In any case $I_{LCS}$ is a topological invariant, $CS(M^3)$,
for a closed oriented Riemann manifold $M^3$, which takes values
on the circle $(\real/\mezzo\zet)\sim S^1$\qquad
\Ref\mayer{
R.Mayerhoff, ``Hyperbolic Three-Manifolds with Equal Volumes but Different
CS-Invariants'', in ``Low Dimensional Topology and Kleinian Groups'', edited
by D.B.A.Epstein, London Math.Soc.Lect.Note Series. 112.}
if $M^3$
is homeomorphic to a closed hyperbolic three-manifold
\foot{According to a famous conjecture
\Ref\thur{
W.B.Thurston, Bull.A.M.S. {\bf 6} (1982) 357.}
about three-manifolds, almost all
interesting (irreducible) three-manifolds have a ``geometrical decomposition''
into  (closed) hyperbolic three-varieties.}.
Therefore, it does not participate to the local short-distance scale
structure of 3D-QG and, hence, to the above computation of UV-divergences.
\chapter{Classical Background Metric from Global Observables of Pure
3D-Quantum Gravity}
In this final part of the paper, we suggest a possible way out of the
conceptual problem raised by Witten
\Ref\wittentre{
E.Witten, Nucl.Phys. {\bf 323B} (1989) 113.}
on how to introduce a background space-time metric as some expectation
value of gauge and diffeomorphism-invariant observables, the Wilson lines,
\Ref\varie{
S.P.Martin, Nucl.Phys. {\bf 327B} (1989) 178;
\nextline
S.Carlip, Nucl.Phys. {\bf 324B} (1989) 106 and Phys.Rev. {\bf 42D} (1990)
2647;
\nextline
J.E.Nelson and T.Regge, Nucl.Phys. {\bf 328B} (1989) 190.}
of the topological pure 3D-Qantum Gravity (QG).
Clearly, this problem underlies the coupling
of matter degrees of freedom to 3D-gravity as discussed, for instance, in
the previous sections. Indeed, the gravitational coupling of
``non-topological''
matter fields only makes sense in the ``broken phase'' of general relativity
where there is a riemannian space-time with distances and light-cones, while
the topological nature of the theory depends upon having
an unbroken phase $\overline {e}^a_\mu =\langle e^a_\mu\rangle=0$
without metric or riemannian interpretation.

Pure 3D-QG has a set of observables, the so-called Rovelli and Smolin [3]
(also Nelson and Regge [15]) observables. One of them will play a
fundamental role in explaining a pure ``quantum''
mechanism that leads to a non-trivial
background space-time metric as an ``order parameter'' for the diffeomorphism
group.
Let us show this mechanism by considering
the pure Euclidean 3D-QG where the Lorentz
group is $SO(3)\sim SU(2)$. Following Rovelli and Smolin, we shall use
for this purpose a part of their observable called ${\cal T}^1$ written here as
${\cal T}^1\equiv Tr{\cal W}_\mu[C](s)$, where
$Tr$ is the trace in the fundamental representation of $SU(2)$.
Naively, one may understand
${\cal W}_\mu[C](s)$ as the parallel
displacement generator
of the dreibein $e^a_\mu$ along a loop $C=C(s)$.
It is defined as follows.
Let us assume for simplicity that the Euclidean space-time $M^3$ is
homeomorphic to $\real^3$. Then, for any loop (knot) $C\in \real^3$ and loop
parameter $s$, ${\cal W}_\mu[C](s)$ is given by inserting $E_\mu(x)\equiv
e^a_\mu(x)\tau_a$, where $\tau_a$ are the generators of $su(2)$, along the
holonomy ${\cal H}(C)$ of $C$ at the point $x=C(s)$, \ie
$$\left\{
\eqalign{
&{\cal W}_\mu[C](s) =E_\mu(C(s)){\cal H}(C)\equiv {\cal W}^a_\mu[C](s)
\tau_a \cr
&{\cal H}(C) \equiv {\cal P}exp(\oint\nolimits_C dx^\mu\ \omega^a_\mu(x)\tau_a)
\cr}\right.. \eqno (18)$$
Here,
${\cal P}$ stands for path-ordering and $\omega^a_\mu$ is the spin-connection,
\ie\ the gauge-connection for the Euclidean Lorentz group. ${\cal T}^1$ is not
reparametrization invariant since it depends on a preferred value of the loop
parameter $s$. However, it is reparametrization covariant in the sense that
${\cal W}_\mu[C'](s)={\cal W}_\mu [C](f(s))$, where $C'(s)=C[f(s)]$,
$f'(s)>0$, is
a reparametrization of $C$ with the same orientation.
Thus, our basic
idea for measuring in a ``gauge invariant way'' the distance between two
points $x$ and $x'$, is to connect them with two half-paths $(\gamma/2)$ and
$(\gamma^{-1}/2)$, together forming a loop $\gamma$ as in
\FIG\loop{The loop $\gamma={\gamma\over 2}\cup {\gamma^{-1}\over 2}$.}
Fig.~\loop, where
$x=\gamma(s)$ and $x'=\gamma(s')$.

Then, we define
\foot{A similar idea of detecting the space-time geometry from the
$ISO(2,1)$-Wilson lines has been recently suggested by S.Carlip.
\Ref\carlip{
S.Carlip, Class.Quan.Grav. {\bf 8} (1991) 5.}}
as a background classical
space-time metric $g^{cl}_{\mu\nu}(x)$, which must be by definition a
c-number, the following
expectation value of the trace of the product of two composite operators
${\cal W}_\mu[\gamma](s)$
evaluated at the points $x=\gamma(s)$ and $x'=\gamma(s')$:
$$g^{cl}_{\mu\nu}(x)=\lim_{{\scriptstyle x'\rightarrow x\atop \scriptstyle
x=\gamma(s)}\atop\scriptstyle x'=\gamma(s')}
\langle Tr({\cal W}^a_\mu[{\gamma\over 2}](s)\tau_a
{\cal W}^c_\nu[{\gamma^{-1}\over 2}](s')\tau_c)
\rangle. \eqno (19)$$
Notice that the operator inside the expectation value of (19) is called by
Rovelli and Smolin ${\cal T}^2$ observable.
The evaluation
of Eq. (19) is quite complicated and for our aim it is sufficient to
limit ourselves to calculating it
in the tree-approximation. Thus, at lowest order in the expansion of the
path ordered ${\cal H}$ in ${\cal W}^a_\mu[\gamma]$ we get
$$-{1\over 4}
\lim_{\varepsilon\rightarrow 0}\oint\nolimits_\gamma dz^\rho\ \langle
e^a_\mu (x_\varepsilon)\omega^b_\rho (z)\rangle
\lim_{\varepsilon\rightarrow 0}\oint\nolimits_\gamma dw^\sigma\ \langle
e^c_\nu (x_\varepsilon)\omega^d_\sigma (w)\rangle Tr(\tau_a\tau_b\tau_c
\tau_d).  \eqno (20)$$
Here, we have used the fact that the trace of an odd product of Pauli matrices
vanishes and that $\langle ee\rangle=0=\langle\omega\omega\rangle$. Notice that
in this tree-approximation and at lowest order in the expansion of the path
ordered, Eq. (20) looks like (up to a numerical factor)
the product of two $\langle {\cal T}^1\rangle$
evaluated in the same approximation.
Furthermore, in order to regularize $\langle e\oint\nolimits_\gamma\omega
\rangle$, we have taken
(see \FIG\twist{An example of framing. The outer curve represents the
framed contour.} Fig.~\twist)
the point $x$, which in the following will be denoted by
$x_\varepsilon$, on the ``framed path'' $\gamma_f$ defined by
$$\gamma_f=\{x^\mu+\varepsilon n^\mu(t)\ :\ |n(t)|=1,\ \varepsilon>0\},
\eqno (21)$$
where $n^\mu$ is a vector orthogonal to $\gamma$ obtained by shifting
the path $\gamma$ on which we calculate the holonomy ${\cal H}(\gamma)$.
In Eq. (20) we shall use the formula
$$ \lim_{\varepsilon\rightarrow 0}\oint\nolimits_\gamma dz^\rho\ \langle
e^a_\mu (x_\varepsilon)\omega^b_\rho (z)\rangle=\lim_{\varepsilon\rightarrow 0}
\varepsilon_{\mu\rho\nu}\partial^\nu_{x_\varepsilon}\oint\nolimits_\gamma
dz^\rho\ {1\over |x_\varepsilon -z|}\delta^{ab}. \eqno (22)$$
The result (22) has been obtained by using
the expansion of $e^a_\mu$ around the ``topological vacuum''
$$e^a_\mu=h^a_\mu,\qquad \overline{e}^a_\mu=\langle e^a_\mu\rangle=0$$
and the form of the propagator $\langle h^a_\mu\omega^b_\rho\rangle$,
Eq. (10.a). Clearly in Eq. (22) there exists an underlying definition of
distance through the modulus. But following Witten [1;14],
we may assume to fix, ``at priori'', an external space-time
metric $\delta_{\mu\nu}$ which will be identified in a ``self-consistent'' way,
in the end, with a flat-order approximation of
$g^{cl}_{\mu\nu,tree}(x)=\delta_{\mu\nu}+x^\rho\partial_\rho
g^{cl}_{\mu\nu,tree}(0)+\ldots$.
In any case the metric dependence enters only the gauge fixing
procedure and does not affect the physical space. An easy way of
understanding this is to recast the first-order dreibein form
of the Euclidean 3D-Einstein action in an $ISO(3)$ Chern-Simons
form [1]. Then one should recognize that the associated symmetric
energy-momentum tensor $T$ is given by the commutator with the BRST charge $Q$,
that is $T=[Q,\ldots]$, where the other member of the commutator is
not relevant here.
\Ref\mgm{
E.Guadagnini, M.Martellini and M.Mintchev, Phys.Lett. {\bf 227B} (1989) 111.}
Since $Q$ annihilates the physical states, the mean value of $T$ vanishes
between physical states.
This implies general covariance on the physical space, as it should be.
Going back to Eq. (22), it is identical, up
to $\delta^{ab}$, to
the potential ${\cal A}_\mu(x_\varepsilon)$ due to a closed magnetic vortex
line
$\gamma$. By the Biot-Savart law, ${\cal A}_\mu(x_\varepsilon)$
can also be interpreted as the
total magnetic field generated by a steady unitary current flowing through
$\gamma$ and observed at a point $x_\varepsilon$ belonging to a curve
$\gamma_f$ that is twisted around $\gamma$ (see Fig.~\twist). The twists
are necessary in order to have a non-trivial result. For instance, if $\gamma$
is a circle and $\gamma_f$ is parametrized (in $\real^3$) by:
$$\gamma_f:\left\{
\eqalign{
&x^1 =(1+\varepsilon \cos\theta)\cos\theta \cr
&x^2 =(1+\varepsilon \sin\theta)\sin\theta \cr
&x^3 =\varepsilon \sin\theta \cr}\right. \eqno (23)$$
setting $\varepsilon_{\mu\nu\rho}=\varepsilon_{123}=1$, we find that
${\cal A}_\mu$ is equal to:
$$\eqalign{
&\lim_{\varepsilon\rightarrow 0} {\cal A}_1(x_\varepsilon)|_{x_\varepsilon=
\gamma_f}=-\lim_{\varepsilon\rightarrow 0}{1\over \varepsilon \cos\theta}
{d\over d\theta}(\cos\theta G_\varepsilon(\theta;\gamma)) \cr
&\lim_{\varepsilon\rightarrow 0} {\cal A}_2(x_\varepsilon)|_{x_\varepsilon=
\gamma_f} =\lim_{\varepsilon\rightarrow 0}{1\over \varepsilon \cos\theta}
{d\over d\theta}(\sin\theta G_\varepsilon(\theta;\gamma)) \cr
&\lim_{\varepsilon\rightarrow 0} {\cal A}_3(x_\varepsilon)|_{x_\varepsilon=
\gamma_f}=\lim_{\varepsilon\rightarrow 0}
[-{1\over \sin\theta(1+2\varepsilon \cos\theta)}{d\over d\theta}
(\cos\theta G_\varepsilon(\theta;\gamma)) \cr
&\qquad\qquad\qquad\qquad\qquad\ +{1\over \cos\theta(1+2\varepsilon
\sin\theta)}{d\over d\theta}(\sin\theta G_\varepsilon(\theta;\gamma))] \cr}.
\eqno (24.a)$$
Here, $G_\varepsilon(\theta;\gamma)$ is defined as
$$\eqalign{
G_\varepsilon(\theta;\gamma)={4\over B(\varepsilon)}&\biggl[
{A(\varepsilon)\over
\sqrt {A(\varepsilon)+B(\varepsilon)}}F\biggl(
{\pi\over 2},\sqrt {{2B(\varepsilon)
\over A(\varepsilon)+B(\varepsilon)}}\biggr) \cr
&- \sqrt {A(\varepsilon)+B(\varepsilon)}
E\biggl({\pi\over 2},\sqrt{{2B(\varepsilon)
\over A(\varepsilon)+B(\varepsilon)}}\biggr)\biggr]
\cr}, \eqno (24.b)$$
where $F$ and $E$ are elliptic integrals of first and second kind and
$A(\varepsilon)$, $B(\varepsilon)$ are suitable functions of $\varepsilon$
such that
$$\lim_{\varepsilon\rightarrow 0} A(\varepsilon)=\lim_{\varepsilon\rightarrow
0}
B(\varepsilon)=2. \eqno (24.c)$$
The same result is obtained from the second factor
of Eq. (20). Thus, we see that $g^{cl}_{\mu\nu,tree}(x)$ is given by the
symmetric $3{\rm x}3$-matrix
$$g^{cl}_{\mu\nu,tree}\simeq const\pmatrix{
{\cal A}_1^2 & {\cal A}_1{\cal A}_2 & {\cal A}_1{\cal A}_3 \cr
\cdots & {\cal A}_2^2 & {\cal A}_2{\cal A}_3 \cr
\cdots & \cdots & {\cal A}_3^2 \cr}, \eqno (25)$$
where in the constant we have absorbed the factor $(1/4)\delta^{ab}\delta^{cd}
Tr(\tau_a\tau_b\tau_c\tau_d)$.
${\cal A}_\mu$, obtained in Eq.'s (24), is non-vanishing, non-topological
and depending on the length cut-off $\varepsilon$, so that it is singular
when this cut-off is removed.
\foot{Notice that $\varepsilon$ can be reabsorbed, after a constant Weyl
transform, by a renormalizaton of the Newton coupling constant.}
It is this regularization, or framing procedure, that breaks the
diff-invariant unbroken phase at the level of the ``true'' ground state of
the ${\cal T}^2$-observable. In other words, roughly speaking, the square root
of the mean field $\langle{\cal T}^2\rangle$
is a collective state which plays the role of a classical background
dreibein field $\overline{e}^a_\mu (x;\varepsilon)=\delta^a_\mu
{\cal A}_\mu (x_\varepsilon)$. On the other
hand, the above framing procedure and hence the $\varepsilon$-dependence,
are strictly necessary in order to get knot (link)-invariant quantities
\Ref\martello{For a perturbative approach as the one followed here see \eg
\nextline
E.Guadagnini, M.Martellini and M.Mintchev, Nucl.Phys. {\bf 330B} (1990) 575.}
starting from the expectation value of the set of R.S.-observables
[3] which are the Wilson line operators
$$
{\cal T}^0 \equiv Tr{\cal H}(C)
\eqno (26)$$
with $C$ a knot (link) in $\real^3$.

We should like to conclude with three remarks.
The first is that the perturbative {\it non}-renormalizability
of 3D-QG in second
order {\it metric}
formalism comes out from the fact that the correlation functions of
the operator-valued metric field are actually expectation values of products
of composite fields (the ${\cal T}^2$-observable). Troubles arise essentially
because we use the Feynman rules for the whole composite fields instead of
those of the fundamental ones (which are $e$ and $\omega$). This situation
perhaps also affects 4D-QG. This understanding of perturbative
non-renormalizability was underlined several times by Ashtekar, Rovelli and
Smolin.

In second place, we have shown
that a classical background Euclidean
space-time metric (Eq. (25)) is already induced at tree-approximation level.
Then, following the common wisdom that regards the tree (semiclassical)
approximation as a large scale (low momenta) limit, we may understand the
$\varepsilon\rightarrow 0$ limit as an infrared one and therefore we may
agree with Witten's claim [14]:

\noindent
``$\ldots$ this infrared divergence is the birth
of macroscopic space-time, starting from microscopic quantum theory''.

Finally, we think that the quantum states of the gravitational field
constructed starting from the loop observables are of two kinds,
``macroscopic'' and ``microscopic''
\Ref\npunto{
N.Seiberg, ``Notes on Quantum Liouville Theory and Quantum Gravity'',
Proc. of the Carg\'ese Meeting
on Random Surfaces, Quantum Gravity and Strings, 1990, RU-90-29},
as it happens in the quantum Liouville approach to 2D-gravity.
The macroscopic states correspond to loop observables like
${\cal T}^0$. The microscopic ones refer instead to operators
like ${\cal T}^1$, where the dreibein explicitely appears.
The former give the global, topological
properties (as generalized knot-link invariants
\Ref\grrr{
B.Br\"ugmann, R.Gambini and J.Pullin, Phys.Rev.Lett. {\bf 68} (1992) 431})
of gravity, while the
latter are associated with the local, metric properties of
the gravitational field.

One of the authors (G.B.) would like to thank G.Grillo for his comments
on the last computational part of this paper.

\refout
\endpage
\figout
\bye